\documentclass[aps,pra,twocolumn,showpacs]{revtex4}
\usepackage{graphicx} 
\usepackage{dcolumn}   
\usepackage{bm}        
\usepackage{amssymb}
\usepackage{amsfonts}   
\usepackage{amsmath} 
\usepackage{amsthm}
\usepackage{color}
\usepackage{marvosym}

\newcommand{\tr}{\text{tr}}

\newcommand{\bra}[1]{\langle #1|}
\newcommand{\ket}[1]{|#1\rangle}

\newcommand{\ketbra}[1]{| #1\rangle \langle #1|}

\newcommand{\be}{\begin{equation}}
\newcommand{\ee}{\end{equation}}
\newcommand{\bea}{\begin{eqnarray}}
\newcommand{\eea}{\end{eqnarray}}

\newcommand{\WW}{\ensuremath{\mathcal{W}}}

\newcommand{\kommentar}[1]{}

\renewcommand{\vr}{\ensuremath{\varrho}}

\newcommand{\forget}[1]{}

\begin{document}

\title{
Multiparticle entanglement as an emergent phenomenon
%Disproving the T\'oth conjecture
}

\date{\today}

\author{Nikolai Miklin}
\author{Tobias Moroder}
\author{Otfried G\"uhne}
\affiliation{Naturwissenschaftlich-Technische Fakult\"at, 
Universit\"at Siegen, Walter-Flex-Str. 3, 57068 Siegen, Germany}

\begin{abstract}
The question whether global entanglement of a multiparticle quantum system
can be inferred from local properties is of great relevance for the theory
of quantum correlations as well as for experimental implementations. We 
present a method to systematically find quantum states, for which the two- 
or three-body marginals do not contain any entanglement, nevertheless, 
the knowledge of these reduced states is sufficient to prove genuine 
multiparticle entanglement of the global state. With this, we show that 
the emergence of global entanglement from separable local quantum states 
occurs frequently and for an arbitrary number of particles. We discuss 
various extensions of the phenomenon and present examples where global 
entanglement can be proven from marginals, even if entanglement cannot 
be localized in the marginals with measurements on the other parties.
\end{abstract}

\pacs{03.65.Ta, 03.65.Ud}
%, 37.10.Ty}
%03.65.Ta: Foundations of quantum mechanics; measurement theory
%03.65.Ud: Entanglement and quantum nonlocality
%(e.g. EPR paradox, Bell's inequalities, GHZ states, etc.)
%42.50.Xa: Optical tests of quantum theory
%37.10.Ty: Ion trapping

\maketitle

\section{Introduction}
 
The relations between global properties of a system and the 
properties of its parts are central for many debates in 
science. In the field of physics these relations are often 
captured by the notion of  {\it complexity} and, although a 
fully developed theory of complexity is still missing, many 
aspects have been discovered. An interesting concept in these 
discussions is the notion of {\it emergence}, meaning that at 
a global scale properties may be present, which are not present
in the parts of the system. A central property of complex {quantum} 
systems is the possibility of being entangled, meaning in the simplest 
case that the wave function of the system does not factorize. Interestingly, 
it was already noted by Erwin Schr\"odinger in his first work on entanglement 
that this notion concerns also the relation between global and local 
properties \cite{erwin}. The study of this relation is known as 
the {marginal problem} or representability problem, 
and, although it has been studied for decades, it remains a key 
problem in quantum mechanics 
\cite{marginalreview}.

With the advent of quantum information processing the question 
to which extent global entanglement properties can be inferred 
from local data can be reformulated in a precise mathematical manner 
and the tools of entanglement theory can be used to tackle it. 
If pure states are considered, then one can clearly 
infer entanglement from local properties since the reduced states of an 
entangled states  are not pure anymore. 
Recently, this approach has been 
refined significantly by characterizing the single-particle reduced states that belong 
to certain classes of multiparticle entanglement \cite{davidscience, kus}. 
Nevertheless, if the purity of the global state is assumed the results cannot be used for any 
experimental implementation, since noise and decoherence always lead 
to mixed states. For multiparticle mixed states, it is clear that
if the marginals contain already entanglement, then the global state
must be entangled, too. Surprisingly, however, there are multiparticle quantum states
where the two-particle marginals are not entangled at all, but nevertheless 
entanglement of the global state follows already from the marginals. This means that global entanglement 
can be proven by looking at separable marginals only and in this way, 
entanglement may appear as an emergent phenomenon. This has first been observed for entanglement in spin models and in the context of spin squeezing, where the task of proving entanglement from two-body marginals arises naturally \cite{gezaspinwitness, SpinSqueez}.
The same observation has been made with the violation of Bell inequalities: here, the marginals 
may be compatible with a local hidden variable model, but such a model 
can be excluded for the global state by considering the marginals only 
\cite{AcinGisin, tura}.

\begin{figure}[t]
\includegraphics[scale=0.25]{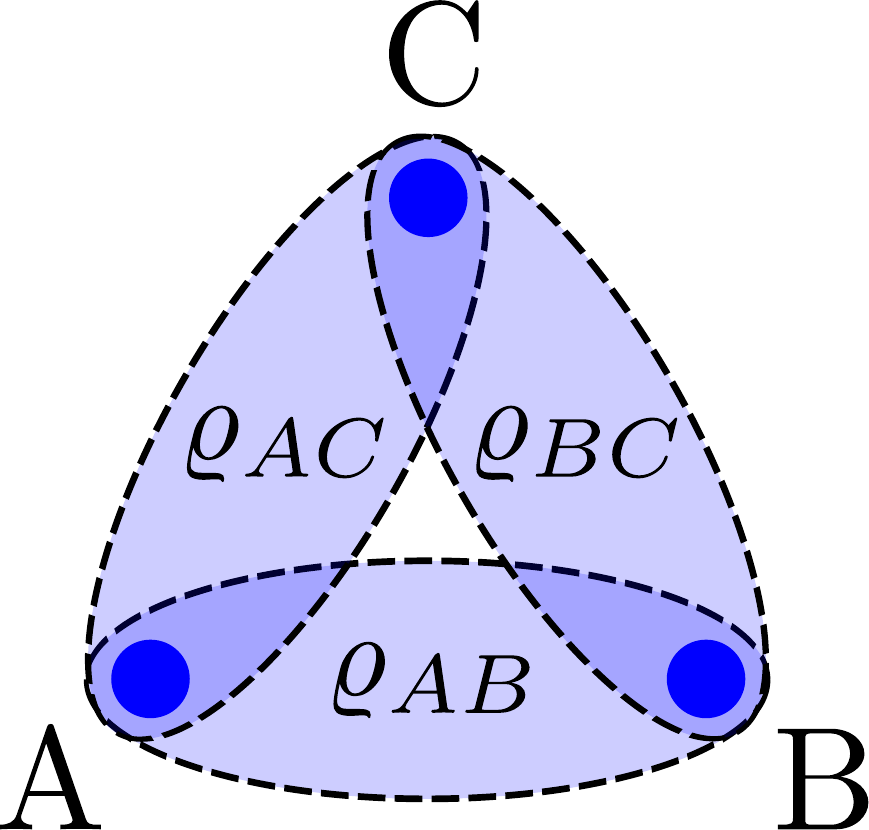}
\hspace{0.5cm}
\mbox{ }
\includegraphics[scale=0.25]{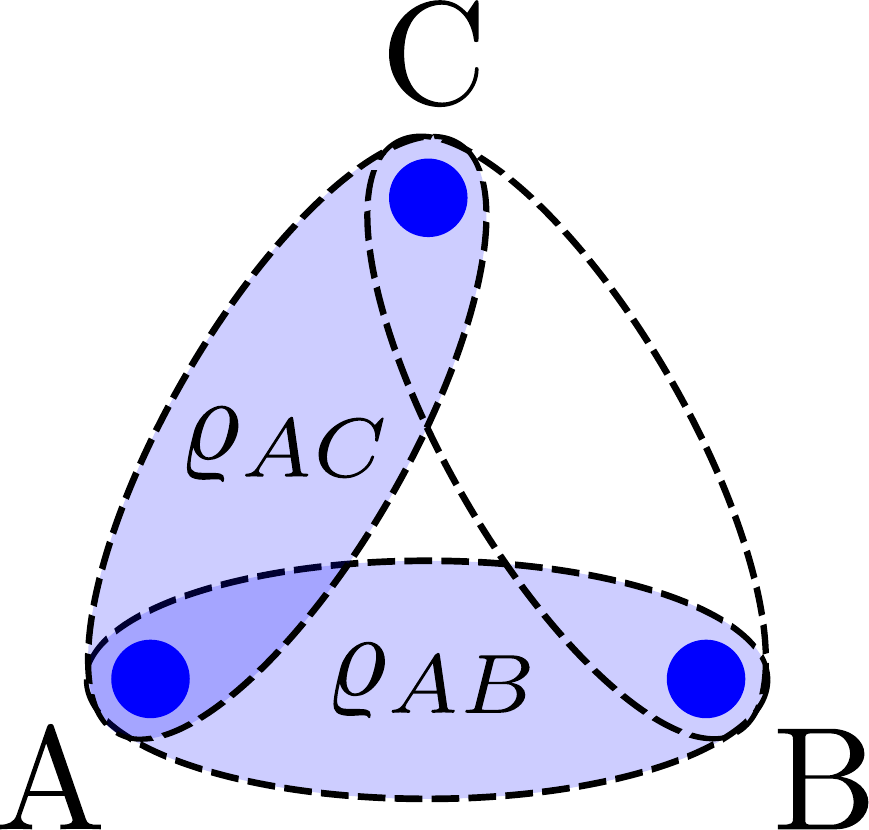}
\hspace{0.5cm}
\mbox{ }
\includegraphics[scale=0.235]{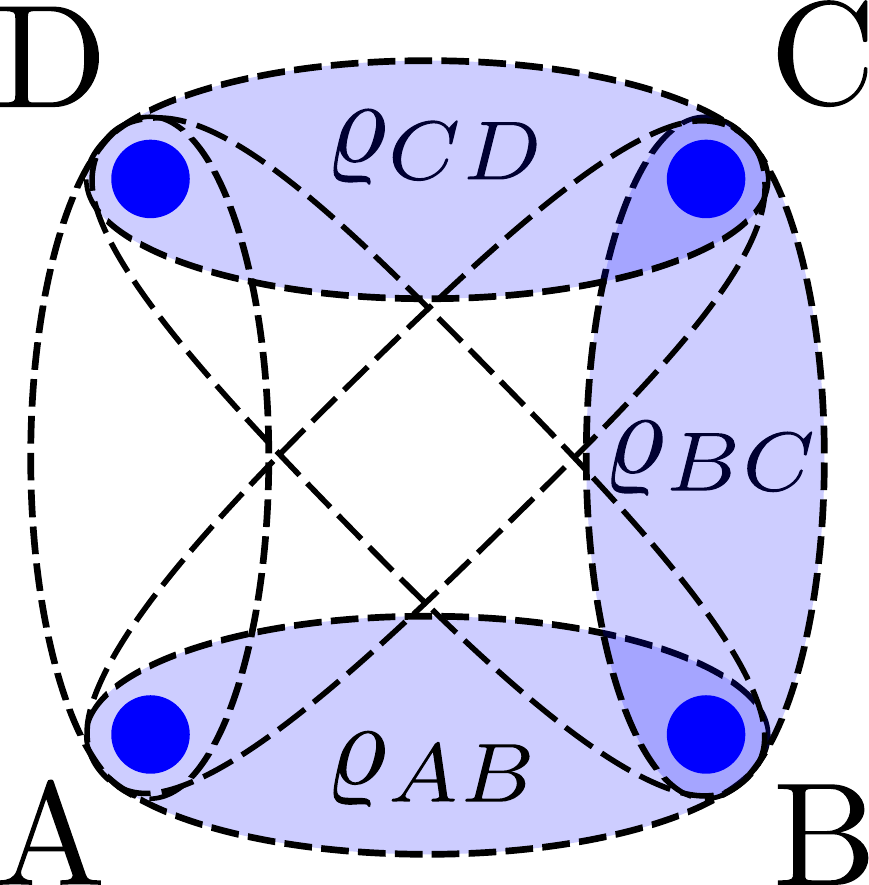}
\caption{ 
Possible applications of method provided in this paper. 
Left: The  method finds systematically three-particle 
states with no entanglement in the two-qubit marginals 
$\vr_{AB}$, $\vr_{BC}$, and $\vr_{AC}$, where nevertheless 
global genuine multiparticle entanglement can be proven 
from knowing the separable marginals only.
Middle, right: Possible variations of the problem considered 
in the paper. All two-body reduced states are required to 
be separable (shown by dashed ellipses), but only some (denoted 
by blue filled ellipses) are  known and sufficient to prove 
that all particles are entangled.  
\label{fig:3q}}
\end{figure}

In all these works, however, only the simplest form of multiparticle 
entanglement was considered. Namely, it was asked whether or not the global 
state factorizes, or in technical terms, 
is fully separable. A state which does not factorize is entangled, 
but this does not mean that all particles are entangled. If all 
particles are entangled the state is called genuine multiparticle
entangled and this property is much more difficult to infer from the 
marginals. In fact, there has been only one recent example of a 
three-qubit state, where the genuine multiparticle entanglement can be
concluded from separable marginals \cite{ChenPiani}. However, this example and the underlying techniques 
are tailored to a specific situation and are not straightforward 
to generalize in any direction.

In this paper, we present a systematic method to find multiparticle
quantum states where genuine multiparticle entanglement can be 
proven from separable marginals (see Fig.~\ref{fig:3q}). Implementing 
this method numerically we find various examples for up to five particles.
Based on these findings we provide a scheme for constructing of states 
with the desired properties for {\it any} number of particles, proving 
the universality of the described phenomenon. Furthermore, we consider possible extensions of the problem. Situations where only a subset 
of the separable two-body marginals is sufficient to prove entanglement, 
can easily be identified, as well as the case, where additionally the separability of higher-order marginals is given. Finally, a direct extension of our method 
delivers states where no entanglement can be generated between two 
arbitrary particles, even if measurements are made on the other 
particles. Still, global entanglement can be proven from the marginals. 
This shows that localizable entanglement is not a precondition for 
genuine multiparticle entanglement. 

\section{Statement of the problem}
Before explaining our method is it useful to recall the main definitions. 
First, a two-particle state $\vr_{AB}$ is separable, if it can be written 
as a convex combination of product states, 
%\be
$
\vr^{\rm sep}_{AB} = \sum_k p_k \vr_A^k \otimes \vr_B^k,
$
%\ee
where the $p_k$ form a probability distribution. Otherwise the state 
is entangled. If three or more particles are considered, different 
kinds of entanglement arise.  To start, a three-particle state is 
fully separable, if it is of the form 
$\vr^{\rm fsep}_{ABC} = \sum_k p_k \vr_A^k \otimes \vr_B^k \otimes \vr_C^k$, otherwise 
it contains some entanglement. But this does not mean that all particle 
are entangled, already the entanglement of two particles suffices to 
exclude full separability. The question whether all particles are 
entangled or not leads to the notion of genuine multiparticle 
entanglement and biseparability. A three-particle state is biseparable, 
if it is of the form
\begin{equation}
\label{gmedef}
\vr_{ABC}^{\rm bisep} = p_1\vr^{\rm sep}_{A|BC} + p_2\vr^{\rm sep}_{B|AC}
+ p_3\vr^{\rm sep}_{C|AB},
\end{equation}
where the state $\vr^{\rm sep}_{A|BC} = \sum_k p_k \vr_A^k \otimes \vr_{BC}^k$ is separable for the partition $A|BC$ etc.
If a state is not biseparable, it is genuine multiparticle entangled.
Clearly, it is more difficult to prove that a state is genuine multiparticle
entangled than proving that it just contains some entanglement,
as more conditions have to be fulfilled.

Now we are able to define the problem in a precise manner. For three 
particles, we want to find a state $\vr$, such that:
{\it (i) All two-body marginals (reduced states) are separable,} 
and
{\it (ii) The state $\vr$ itself is genuine multiparticle entangled 
and the entanglement can be proved from the two-body marginals.} 

In other words, the two conditions state that the separable 
marginals are only compatible with global genuine 
multipartite entanglement. It is important to note that in our problem uniqueness of the global state is not required and the only condition is that all global states, compatible with the marginals, must be genuine multiparticle entangled. One can extend the conditions in 
various ways. Concerning the first condition, one can require 
for four or more particles in addition to the two-body marginals 
that also the three-body marginals are fully separable. The second 
condition can be modified such that only some of the two-body 
marginals are known, while all of them are still separable (see Fig.~\ref{fig:3q}). Clearly, 
both modifications make it more difficult to find the desired 
states and it is not clear, whether states with these properties exist at all. It is one of the main results of the present paper that for all reasonable modifications of 
the problem the corresponding states can be found.

\section{Construction method} 
Our method for constructing states with the desired properties 
is formulated as an iteration process consisting of two optimization 
problems, which are semidefinite programs (SDPs) and can thus be
solved efficiently with freely available tools \cite{YALMIP, SDPT3}.
The algorithm is based on the approach to multiparticle entanglement 
with PPT mixtures \cite{PPTMixer}. This approach verifies 
whether or not a given state $\vr$ is a mixture of states
which have a positive partial transpose (PPT) for the different 
bipartitions
\footnote{The partial transpose of a state 
$\vr= \sum_{ij,kl} \vr_{ij,kl} \ket{i}\bra{j}\otimes\ket{k}\bra{l}$
is given by 
$\vr^{T_A}= \sum_{ij,kl} \vr_{ij,kl} \ket{j}\bra{i}\otimes\ket{k}\bra{l}$
and for a separable state the partial transpose is positive semidefinite.}.
So the test checks whether 
\begin{equation}
\label{PPTdef}
\vr_{ABC}^{\rm pptmix} = p_1\vr^{\rm ppt}_{A|BC} + p_2\vr^{\rm ppt}_{B|AC} + p_3\vr^{\rm ppt}_{C|AB}.
\end{equation}
Since separable states are also PPT, any state that fulfills the biseparability 
condition from Eq.~(\ref{gmedef}) will also be a PPT mixture according to 
Eq.~(\ref{PPTdef}). Consequently a state that is not a PPT mixture must be 
genuine multiparticle entangled. The question whether a state is a PPT mixture 
can directly be decided via SDPs, most importantly, it can also be decided when 
only partial information on the state is available (see also below).

Now we describe the steps of the searching algorithm.  
We formulate it first for qubits, the extension to higher-dimensional systems
is explained afterwards.

{\bf Step 1.} In the first part we apply the criterion for PPT mixtures to 
a random initial state $\vr_0$ using only its marginals. For that, we have to solve the SDP
\begin{eqnarray}
\label{AlgStep1}
\underset{\WW ,P_M,Q_M}{\text{minimize:}} & & \tr(\WW\vr_0),  \\
\text{subject to:} & & \tr(\WW) = 1, \; \text{where} \nonumber \\ 
& & \WW = \sum\limits_{ij} w^{\alpha,\beta}_{ij}\sigma^\alpha_i\otimes\sigma^\beta_j\otimes\openone^{\otimes (N-2)} + \text{perm}.
\nonumber \\
& & \text{and for all bipartitions} \; M|\overline{M}: 
\nonumber \\ 
& & \WW = P_M + Q^{T_M}_M, \; Q_M \geq 0, \; P_M \geq 0. \nonumber
\end{eqnarray}
This optimization program means that for the state $\vr_0$ 
we construct the optimal $\WW$, which is a so-called
decomposable witness for all bipartitions $M|\overline{M}$. 
If the expectation value $\tr(\WW\vr_0)$ for such a witness 
is negative, the state is genuine multipartite entangled, 
more details can be found 
in Ref.~\cite{PPTMixer}.  The first constraint on the witness' 
trace maximizes the white-noise tolerance of the 
entanglement detection. The second linear constraint 
requires the witness to contain only two-body correlations, 
since we want to certify entanglement of our states from 
their two-body marginals only. Permutations are taken for 
all pairs $\alpha,\beta$ of qubits. This constraint is 
additional to the original PPT mixture problem and it is 
one of the algorithm's main features.

{\bf Step 2.} In the second part of the iteration we 
determine the state that gives the most negative
value for the given witness from the first step. In line with
our approach, we require separability of the 
reduced two-party marginals. For the case of qubits this part 
can be formulated as a simple SDP, since 
a two-qubit state is separable, if and only if it is PPT. So we consider:
\begin{eqnarray}
\label{AlgStep2}
\underset{\vr}{\text{minimize:}} & & \tr(\WW\vr) \\
\text{subject to:} & & \vr \geq 0, \; \tr(\vr)=1, \nonumber \\
& & \text{and for all } \alpha,\beta \;\; \vr^{T_\alpha}_{\alpha\beta} \geq 0, \nonumber
\end{eqnarray}
where $\alpha,\beta$ denote pairs of qubits and 
$\vr_{\alpha \beta}$ is the two-qubit marginal. 

Combining both steps 1 and 2 and putting the output $\vr$ of 
the second part as an input into the first part one can run an 
iteration process. If, at some point of the iteration the second 
step gives a negative value, we found already a state that has 
separable two-body marginals, where the entanglement can be 
proven from the marginals only. In practice, if a pure random 
state is taken as a seed of the algorithm, the iteration process 
typically gives a state with the desired properties after the 
second or the third iteration. Running the iteration further 
maximizes the entanglement and noise robustness of the state, 
while keeping the desired properties. Besides, 
the output states for different inputs $\vr_0$ mainly 
differ only up to some local unitary transformation.  

The algorithm can be extended to modifications of the problem 
in several ways. First, if higher-dimensional systems are considered, 
one has to use the appropriate generalizations of the Pauli matrices in the construction of the witness in Eq.~(\ref{AlgStep1}).
Also, in higher dimensions the PPT criterion is not necessary for separability, 
so for the state resulting in the second step separability of the marginals is 
not guaranteed. Nevertheless, if a state is found, the separability can later 
be checked with existing effective algorithms for proving separability of 
quantum states \cite{julio, kampermann}. Second, if only some of the marginals 
are known [as in Fig.~(\ref{fig:3q})] one just has to modify the 
definition of the witness in Eq.~(\ref{AlgStep1}) and take only correlations 
from the known marginals. Finally, in the case of more than three parties, 
if also the full separability of the three-body marginals is required, one 
can modify the second step by changing the conditions that now the three-body marginals are PPT for any bipartition, and later verify the full separability 
of them with existing approaches \cite{julio, kampermann}. 
\section{Results}
\subsection{Three qubits}
Although a state with the desired properties for 
three qubits is already known (see Ref.~\cite{ChenPiani} and Appendix A), 
we start with this case, as this allows to explain our methods.

Taking the most robust algorithm's final state, applying  
local unitary operations and searching for an analytical 
expression, we find the following state which has separable 
two-body marginals which suffice to prove genuine multiparticle 
entanglement:
\begin{eqnarray}
\label{BS3q}
{\vr_N^{(3)}} &=& \frac{2}{3}|\xi\rangle\langle\xi|+\frac{1}{3}|\bar{W_3}\rangle\langle \bar{W_3}|, 
\\ \quad
|\xi\rangle &=& \sqrt{\frac{1}{3}}|W^{\mbox{\tiny \Radioactivity}}_3\rangle + \sqrt{\frac{2}{3}}|111\rangle, \nonumber
\end{eqnarray}
where $|\bar{W_3}\rangle = (|011\rangle + |101\rangle + |110\rangle)/{\sqrt{3}}$
and 
$|W^{\mbox{\tiny \Radioactivity}}_3\rangle =
(e^{i\frac{\pi}{3}}|001\rangle + e^{-i\frac{\pi}{3}}|010\rangle -|100\rangle)
/{\sqrt{3}}$ is a $W$-state with equally distributed phases. 
Note that besides these phases 
the state itself would be permutationally invariant. However, the asymmetry is necessary, since it can be shown that 
for symmetric states the studied phenomenon cannot exist 
\cite{SpinSqueez} \footnote{Nevertheless, we found with our algorithm permutationally 
invariant states with the desired properties.}. We add that one can
also see the set of marginals $\vr_{AB}, \vr_{BC}$ and $\vr_{AC}$ as the output
of the algorithm, we will discuss below the extent to which the marginals determine
the state completely. 

To compare our result with the result from Ref.~\cite{ChenPiani}, 
we note that the phenomenon described here is not fragile,
as can be seen by the white noise tolerance. We consider 
the target state mixed with white noise
$
\vr(p) = (1-p) {\vr_N^{(3)}}+ p {\openone}/{8}
$
and ask, for which values of $p$ the state has still the desired properties. 
Clearly, the marginals remain separable if white noise is added, so one only
has to check whether the three-qubit entanglement can be proven from
the marginals. One finds that the state in Eq.~(\ref{BS3q}) remains this property with possible 13.7\% of white-noise added, which means that this phenomenon is quite robust and it is realistic that this phenomenon can be observed experimentally. In addition, the white-noise tolerance of the state itself is 28.6\%, but then the 
entanglement cannot be proven from the marginals only. For comparison, the state from Ref.~\cite{ChenPiani} keeps its properties only up to 5.2\% of possible white-noise. Note that such estimates are impossible with the methods 
from Ref.~\cite{ChenPiani}, this it is another advantage of our
algorithm.   

\subsection{Four and five qubits}
For four qubits we find many 
analytical examples of states that have the desired properties. 
Remarkably, there are now also pure states with separable marginals, 
from which entanglement can be proved. One of the simplest 
solutions for four qubits is a Dicke-type state without one term 
and with one $\pi$--phase: 
\be
\label{BS4q1}
|N^{(4)}\rangle =\frac{1}{\sqrt{5}}\big(|0011\rangle+|0101\rangle+|0110\rangle  
+|1001\rangle-|1010\rangle\big). 
\ee
After local unitaries this state may also be expressed as a cluster state 
with an extra term $|N^{(4)}\rangle = 2/\sqrt{5}|CL_4\rangle + 1/\sqrt{5}|0110\rangle$. 
This state has 21.2\% white-noise tolerance (here and later we mean by white-noise 
tolerance the tolerance of the studied properties). 
A further property is that the state $|N^{(4)}\rangle$
is uniquely determined by its two-body marginals, this fact has far-reaching
consequences as we will see below.
There are other four-qubit states with the desired properties. The most 
entangled state, which we found for four-qubits is given in Appendix B 
and in Appendix C    we provide a five-qubit example.

\begin{figure}
\begin{center}
\includegraphics[scale=0.23]{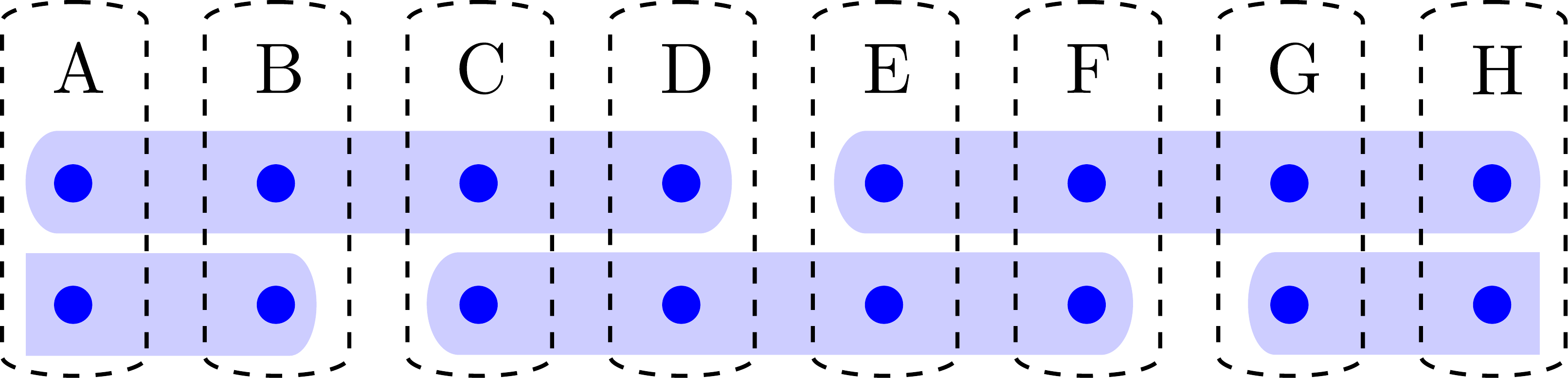}
\end{center}
\caption{
\label{fig:N} 
Illustration of the construction of the desired state for eight parties. Dashed 
lines represent different parties, each party possesses two qubits. Blue rounded 
rectangles depict the entangled four-qubit state $\ket{N^{(4)}}$ from
Eq.~(\ref{BS4q1}).
}
\end{figure}

\subsection{Generalization to more particles} 
Using the state in Eq.~(\ref{BS4q1}) one can construct examples of 
states for an arbitrary number of particles. We can formulate:

\noindent
{\bf Observation.}
{\it
For any number of particles greater than four one can find a pure state with separable
two-body marginals, where genuine multiparticle entanglement can be proven from
these marginals only. 
}

We illustrate the construction for eight parties, where each of the 
parties possesses a four-level system, represented by two qubits, 
so the total system consists of 16 qubits. We can distribute four 
copies of the state $\ket{N^{(4)}}$  in Eq.~(\ref{BS4q1}) as depicted
in Fig.~(\ref{fig:N}). Now every two-party 
marginal is separable since it is a direct product of two two-qubit separable 
states. Furthermore, knowledge of the two-party marginals implies knowledge
on the states $\ket{N^{(4)}}$  and their distribution among the 16 qubits, 
since $\ket{N^{(4)}}$ is uniquely determined by its marginals. 
So the state of eight parties is uniquely determined by its 
two-body marginals as well. The global state is pure and does not factorize for
any bipartition of the eight parties, so it is genuine multiparticle entangled
and the entanglement can be proven from the separable marginals.

A similar reasoning is valid for any number of parties. For five 
parties one takes two copies of the state $\ket{N^{(4)}}$ and distributes the
first copy to the parties $A,B,C,D$ and the second copy to the parties 
$B,C,D,E$, for other numbers of qubits the construction is analogous.
Note that the purity of the state $\ket{N^{(4)}}$ is essential for the argument,
so the construction would not work with other states, e.g. the state from Ref.~\cite{ChenPiani}. Also, we needed that the state is uniquely determined 
by its two-qubit reduced states. Below we discuss this property for the other states we found using the  method.

\subsection{Uniqueness of the global state}
The property of uniqueness of the global sate was noted before for the state from Ref.~\cite{ChenPiani}. In fact, this is the case for both three and four-qubit 
states from the Eqs.~(\ref{BS3q}, \ref{BS4q1}). However, this property does 
not necessarily follow from the constraints and a counterexample
for four qubits is the following state:
\be
\label{BS4q3}
|\Psi\rangle = \frac{1}{\sqrt{2}}|\widetilde{D^4_2}\rangle + \frac{1}{\sqrt{2}}|GHZ_4\rangle
\ee
with
\begin{eqnarray}
|GHZ_4\rangle = \frac{1}{\sqrt{2}}\Big(|0000\rangle + |1111\rangle\Big),\nonumber \\
|\widetilde{D^4_2}\rangle =
\frac{1}{\sqrt{6}}\Big(|0011\rangle + |0101\rangle + |0110\rangle \nonumber \\
+ e^{i\varphi}|1001\rangle + e^{i\varphi}|1010\rangle + e^{-i\varphi}|1100\rangle\Big), \nonumber
\end{eqnarray}
and $\varphi = \arccos(-1/3)$.

In this case, the two-qubit marginals are also compatible with the 
state of the same form as Eq.~(\ref{BS4q3}), but with the opposite 
phases $\varphi \to - \varphi$, meaning that there is a set of 
states, which are convex combinations of $|\Psi(\varphi)\rangle$ 
and $|\Psi(-\varphi)\rangle$ and which are compatible with the same 
reduced two-qubit states. This demonstrates that in our problem the 
set of two-body reduced states need not be compatible with only 
one global state, the only condition is that set of global states, 
compatible with these marginals, must be enclosed into the set of 
genuine multiparticle entangled states. 

\section{Extensions of the problem}
First, one may ask whether there is any four-qubit 
state with separable two {\it and} three-body marginals, where 
the genuine multiparticle entanglement can be proven from the 
two-qubit reduced states. Our approach can also be used to find such
states, details are given in Appendix D 
Second, there may be the possibility to detect entanglement from only 
a part of two-body correlations (see Fig.~\ref{fig:3q}). For the 
three-qubit configuration where only the correlations between Alice 
and Bob and Alice and Charlie are known, we present a state in 
Appendix E. For the four-qubit case we note that
the genuine multiparticle entanglement of the state from the 
Eq.~(\ref{BS4q3}) can be detected by knowing only 
$\vr_{AB},\vr_{BC},$ and $\vr_{CD}$ with possible 3\% of 
white noise added. As a further extension, we found also 
examples for three three-dimensional systems, see 
Appendix F  for details.
Judging by the simulations we made, it seems that it is possible
to find a state for any imaginable configuration of measurable 
correlations as soon as they contain correlations between every 
bipartition of the system.

%\subsection*{Genuine entangled states with no entanglement in the subsystems} 
Finally, we demonstrate that an extension of our method can be used
to find a three-qubit state, where the marginals do not contain any
entanglement, even after arbitrary measurements on the third particle. 
Nevertheless, global entanglement can be proved from the marginals 
only.

\begin{figure}[t]
\begin{center}
\includegraphics[scale=0.30]{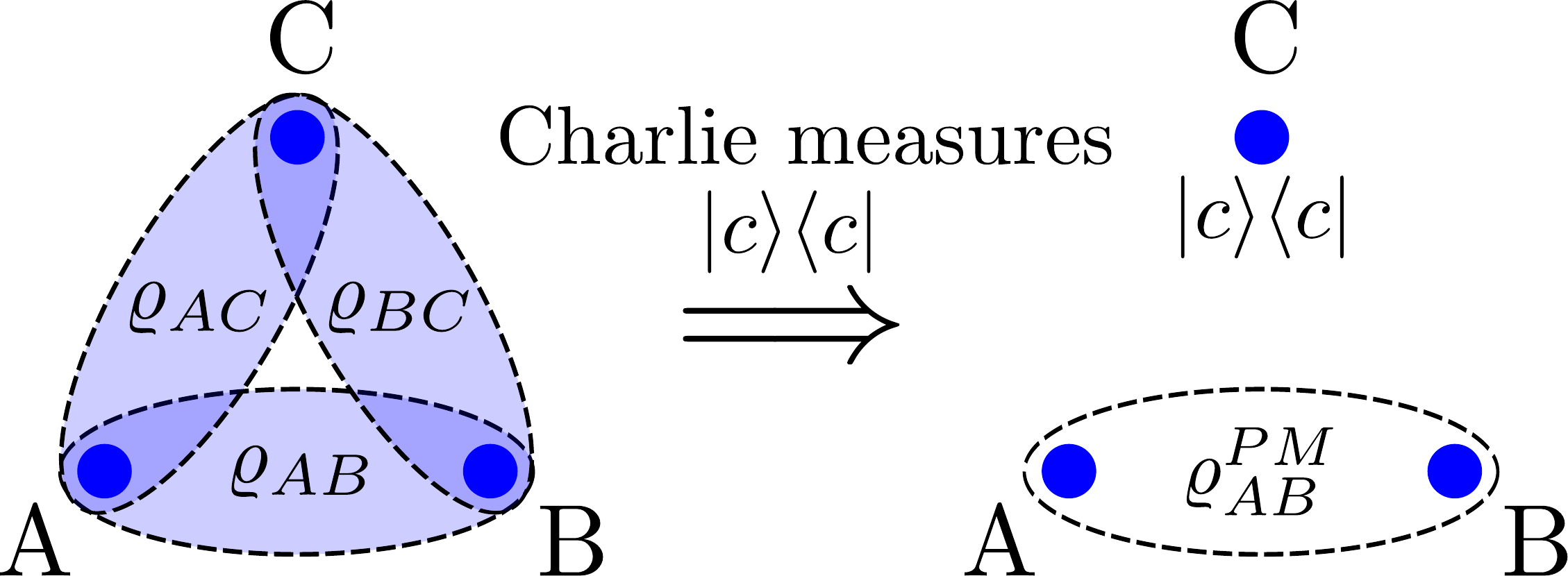}
\end{center}
\caption{
\label{fig:M}
Illustration of the emergence of multiparticle entanglement
in the case of vanishing localizable entanglement. The 
post-measurement state $\vr_{AB}^{PM} \sim \bra{c}\vr_{ABC}\ket{c}$
is separable after arbitrary
measurements of Charlie, but knowing the reduced states $\vr_{AB},
\vr_{AC},$ and $\vr_{BC}$ is still enough to prove genuine tripartite 
entanglement of the original state $\vr_{ABC}$.
}
\end{figure}

To start, one can ask whether the two-particle marginals are also 
separable, if the third party has made a measurement, described
by a projector $\ketbra{c}$ (see Fig.~\ref{fig:M}). For projective 
measurements, the post-measurement state is described by 
$\vr_{AB}^{PM} \sim \bra{c}\vr_{ABC}\ket{c}$ and one can see 
that if this state is separable for all $\ket{c}$ then the marginal 
is also separable, if Charlie makes a generalized measurement or an operation
\footnote{The most general marginal after an operation on $C$ is given 
by $\vr_{AB}^{PM} = 
{\rm Tr}_C[\mathbb{I}_A\otimes\mathbb{I}_B \otimes \Lambda_C(\vr_{ABC})]$,
where $\Lambda_C(\rho)= \sum_k A_k \rho A_k^\dagger$ with 
$\sum_k A_k^\dagger A_k \leq \openone$ is a trace-non-increasing 
operation. Using the polar decomposition 
$A_k=U [s_1\ketbra{c}+s_2\ketbra{c^\perp}]$ with $s_i \geq 0$ 
we have for a single term that ${\rm Tr}_C[A_k (\vr_{ABC})A_k^\dagger]
=
s_1\bra{c}\vr_{ABC}\ket{c} + s_2\bra{c^\perp}\vr_{ABC}\ket{c^\perp}$
is separable as a convex combination of separable states, so also $\vr_{AB}^{PM}$ is separable.}.
So the question arises whether there exist genuine tripartite 
entangled states with the property that whatever projective measurement 
one party performs, the remaining two parties share a separable 
state. One can think about it as a state with no creatable entanglement 
in its subsystems, or a state with vanishing localizable entanglement 
\cite{LocEnt}. In addition, we want the entanglement to be provable
from the two-body marginals $\vr_{ij}$, but the knowledge of the 
conditional states $\vr_{ij}^{PM}$ should not be required 
for the entanglement proof.

To look for such a state one should add an infinite number 
of additional constraints to the semidefinite program,
corresponding to all possible local projective measurements. 
This is, of course, infeasible. However, it is sufficient to 
add a {\it finite} number of constraints corresponding to some 
measurements (preferably equally distributed over the Bloch sphere) 
and require a strict positivity (e.g., $[\vr_{AB}^{PM}]^{T_A} 
\geq \varepsilon \openone$ for some $\epsilon >0$) of the partial 
transposition of the post-measurement states. If a state is finally 
found, one can check by direct numerical evaluation whether the 
post-measurement states are separable even for arbitrary measurements.

Applying this in practice, one directly finds states with the desired 
properties, an example is given in detail in Appendix G. This proves 
that genuine multiparticle entanglement can emerge, even
if the  localizable entanglement vanishes.

\section{Conclusions} 
In conclusion, we have provided a method to study systematically the 
emergence of genuine multiparticle entanglement from separable marginals. 
Our findings show the rich structure of multiparticle entanglement, where
essentially all possible entanglement properties of the marginals can be
combined. We believe that the entanglement properties of the resulting 
quantum states deserve further study, moreover, it would be interesting
to observe the here described effects experimentally.

\begin{acknowledgments}
We thank O. Gittsovich, C. Schwemmer, J. Siewert, C. Spee, G. T\'oth, R. Uola and
an anonymous referee for comments and discussions.
This work has been supported by 
the DAAD, 
the EU (Marie Curie CIG 293993/ENFOQI),
the ERC (Consolidator Grant 683107/TempoQ)
the FQXi Fund (Silicon Valley Community Foundation) 
and the DFG.
\end{acknowledgments}

\section*{Appendix}

\subsection{The state found in Ref.~[9]}
In Ref.~[9] the authors provide an example of 
two-qubit marginal triplet, which is compatible with only 
one global state, which is genuine tripartite entangled. It is
given by:
\begin{eqnarray}
\label{ChenPianiSt}
{\varrho} &=& \frac{2}{3}|\xi\rangle\langle\xi|+\frac{1}{3}|111\rangle\langle 111|, \\ 
|\xi\rangle &=& \frac{1}{2}|010\rangle + \frac{1}{2}|100\rangle + \frac{1}{\sqrt{2}}|001\rangle. \nonumber
\end{eqnarray}

\subsection{The four-qubit state with the highest noise robustness}
The most noise-robust four-qubit state that we found contains a Dicke 
part and a GHZ part with the Dicke part having asymmetric amplitudes.  
It is given by:
\begin{eqnarray}
\label{BS4q4}
|\Psi\rangle &=& \frac{1}{\sqrt{2}}|\eta \rangle + \frac{1}{\sqrt{2}}|GHZ_4\rangle,\\
|\eta \rangle &=& \frac{1}{2\sqrt{2}}\big(-i|0011\rangle + |0101\rangle + |0110\rangle \nonumber \\ 
&-& \sqrt{3}|1001\rangle + i|1010\rangle + |1100\rangle \big). \nonumber
\end{eqnarray}
This state has white-noise tolerance of 22.4\% and it is the closest one to the numerical solution.

\subsection{A five-qubit example}
The following state is a genuine multiparticle entangled five-qubit state 
with separable marginals, which are compatible with this state only. 
\begin{eqnarray}
\label{BS5q}
&&|\Psi\rangle = \frac{1}{\sqrt{6}}|00\rangle\otimes|000\rangle \\
&+& \frac{1}{\sqrt{8}}|11\rangle\otimes\big(\sqrt{\frac{2}{3}}|000\rangle-|001\rangle-|010\rangle-|100\rangle\big) \nonumber \\
&+& \frac{1}{\sqrt{24}}|01\rangle\otimes\big(-|001\rangle + e^{-i\frac{\pi}{3}}|010\rangle+e^{i\frac{\pi}{3}}|100\rangle\big) \nonumber \\
&+& \frac{1}{\sqrt{48}}|01\rangle\otimes\big(e^{-i\frac{2\pi}{3}}|011\rangle + e^{i\frac{2\pi}{3}}|101\rangle+|110\rangle\big) \nonumber \\
&+& \frac{1}{\sqrt{24}}|10\rangle\otimes\big(-|001\rangle + e^{i\frac{\pi}{3}}|010\rangle+e^{-i\frac{\pi}{3}}|100\rangle\big) \nonumber \\
&+& \frac{1}{\sqrt{48}}|10\rangle\otimes\big(e^{-i\frac{\pi}{3}}|011\rangle + e^{i\frac{\pi}{3}}|101\rangle-|110\rangle\big). \nonumber
\end{eqnarray} 
This state has 17.3\% white-noise tolerance.

\subsection{A four-qubit state with separable two and three-body marginals}

To find the following example, we have used our method to find a state, where the 
two- and three-body marginals are all PPT, and still the entanglement can
be proven from the two-body marginals. Our method results in the following
state
\begin{eqnarray}
\label{BS4q2}
\vr_{N}^{(4)} &=& \frac{1}{2}|\zeta_1\rangle\langle\zeta_1| + \frac{1}{2}|\zeta_2\rangle\langle\zeta_2|, \\ 
|\zeta_1\rangle &=& \sqrt{\frac{4}{5}}|GHZ_4\rangle + \sqrt{\frac{1}{5}}|\Psi^+\rangle_{AB}\otimes|\Psi^+\rangle_{CD}, \nonumber \\
|\zeta_2\rangle &=& \sqrt{\frac{2}{5}}\big(|0011\rangle+|1100\rangle\big) + \sqrt{\frac{1}{5}}|\Psi^-\rangle_{AB}\otimes|\Psi^-\rangle_{CD}, \nonumber
\end{eqnarray}
where $|\Psi^\pm\rangle = (|01\rangle \pm |10\rangle)/\sqrt{2}$ 
denote the Bell states. This state has the described properties, even 
if up to 21.8\% of white-noise are added. However, while the PPT property 
of the two-body marginals implies separability this is not the case for 
the three-body marginals: Even if a three-qubit state is PPT for all 
bipartitions, this does not mean that it is fully separable. Therefore, 
we checked whether the three-qubit marginals are fully separable with 
the separability testing algorithm from Ref.~[13] and found that 
if more than 13.5 \% of white noise is added, the marginals are indeed 
fully separable. This proves that the state 
$\vr(p) = (1-p) {\vr_N^{(4)}}+ p {\openone}/{16}$ for $p\in [0.135, 0.218]$
has the desired properties.

\subsection{Three-qubit case, where not all marginals are known}
A genuine multiparticle entangled state with all two-body marginals 
being separable and which can be detected from the correlations 
between A and B and A and C  only (as depicted in 
Fig. 1) is given by: 
\begin{eqnarray}
\label{BS3q2}
{\vr} &=& \frac{1}{{2}}|\xi_1\rangle\langle\xi_1|+\frac{1}{{2}}|\xi_2\rangle\langle\xi_2|, \\
|\xi_1\rangle &=& \sqrt{\frac{1}{10}}\big(\sqrt{5}|000\rangle + \sqrt{4}\,e^{-i\frac{3}{4}\pi}|011\rangle + e^{-i\frac{3}{4}\pi}|101\rangle\big), \nonumber \\
|\xi_2\rangle &=& \sqrt{\frac{1}{10}}\big\{\sqrt{3}\big(|001\rangle + e^{i\frac{2}{3}\pi}|010\rangle + e^{-i\frac{1}{3}\pi}|100\rangle\big) + |111\rangle\big\}. \nonumber
\end{eqnarray}  
This effect has 5\% of white-noise tolerance.

\subsection{A three-qutrit example}

When considering higher dimensions, it is worth 
noting that for every example of states, which 
we present in this paper, there exist trivial 
extensions to higher dimensions. For instance, 
if we take a three-qubit state $\vr_N^{(3)}$ from 
Eq.~(5), then, as noted in Ref.~[9], 
a family of states 
$\vr^{(3)}_d = p_1\vr_N^{(3)} 
+ \sum^d_{m=2}p_m|mmm\rangle\langle mmm| \;(p_1>0, p_m\geq 0)$ 
satisfy all the conditions of the desired states. 
However, this extension is rather trivial and does not give 
the best possible effect for qudit states. Using our algorithm for 
the three-qutrit case, we have found the following state. 
\begin{eqnarray}
\label{BS3qt}
\vr &=& \frac{1}{2}|\eta_1\rangle\langle\eta_1| + \frac{1}{2}|\eta_2\rangle\langle\eta_2|, \\
|\eta_1\rangle &=& \frac{1}{\sqrt{12}}(|000\rangle-|222\rangle) - \frac{i\sqrt{5}}{6}(|012\rangle + |021\rangle \nonumber \\
&-& |102\rangle + |120\rangle + |201\rangle + |210\rangle), \nonumber \\
|\eta_2\rangle &=& \frac{1}{\sqrt{6}}|111\rangle - \frac{\sqrt{5}}{6}(|012\rangle - |021\rangle + |102\rangle \nonumber \\
&+& |120\rangle + |201\rangle - |210\rangle). \nonumber
\end{eqnarray}
for which the two-body marginals are PPT, but genuine multiparticle
entanglement can be proven from them. The white-noise tolerance of 
the discussed properties of the state is 29.5\%. Separability of 
its reduced states can be proven using the algorithm [13]
if 5.3\% of white-noise is added. So, the three-qutrit state 
$\sigma=(1-p)\vr+p \openone/27$ has for $p \in [0.053,0.275]$ the 
property that genuine multiparticle entanglement can be proven from
separable marginals.

\subsection{A three-qubit state without entanglement in the marginals after measurements}
The following state is a three-qubit state with separable marginals 
after an arbitrary projective measurement on one of the qubits but where
the two-body correlations are sufficient to determine genuine 
multiparticle entanglement of the global state. 
\begin{eqnarray}
\label{BS3qM2}
\vr &=& \frac{1}{3}|\chi_1\rangle\langle\chi_1| + \frac{1}{3}|\chi_2\rangle\langle\chi_2| \\ 
&+& \frac{1}{6}|\chi_3\rangle\langle\chi_3| + \frac{1}{6}|\chi_4\rangle\langle\chi_4|, \nonumber 
\end{eqnarray}
where the eigenvectors are
\begin{eqnarray}
\label{BS3qM2chi}
|\chi_1\rangle &=& \sqrt{\frac{5}{21}}\big(|001\rangle + e^{-i\frac{\pi}{6}}|010\rangle + e^{-i\frac{3}{4}\pi}|100\rangle\big) \\
&+& \sqrt{\frac{2}{21}}\big(e^{i\frac{\pi}{5}}|011\rangle + e^{i\pi}|101\rangle + e^{i\frac{\pi}{9}}|110\rangle\big), \nonumber \\
|\chi_2\rangle &=& \frac{1}{3}e^{i\frac{4}{5}\pi}|000\rangle + \sqrt{\frac{3}{7}}|111\rangle \nonumber \\
&+& \sqrt{\frac{1}{42}}\big(e^{i\frac{5}{6}\pi}|001\rangle + e^{-i\frac{2}{3}\pi}|010\rangle + e^{-i\frac{3}{5}\pi}|100\rangle\big) \nonumber \\
&+& \sqrt{\frac{7}{54}}\big(e^{-i\frac{3}{5}\pi}|011\rangle + e^{-i\frac{5}{9}\pi}|101\rangle + |110\rangle\big), \nonumber \\ \nonumber \\
|\chi_3\rangle &=& \frac{\sqrt{18}}{5}|000\rangle + \frac{1}{5}e^{i\frac{\pi}{5}}|111\rangle \nonumber \\
&+& \frac{\sqrt{2}}{5}\big(e^{i\pi}|001\rangle + e^{-i\frac{\pi}{2}}|010\rangle + e^{-i\frac{2}{5}\pi}|100\rangle\big), \nonumber \\
|\chi_4\rangle &=& \frac{1}{\sqrt{3}}\big(|001\rangle + e^{-i\frac{5}{6}\pi}|010\rangle + |100\rangle\big). \nonumber
\end{eqnarray}
This state keeps its properties with possible level of white noise of 2\%. As discussed in the corresponding section, to find this state we added a finite number of constraints, corresponding to various projective measurements, and then required strict positivity of the post-measurement states. In our implementation we defined $\sim 1000$ constraints and required the eigenvalues of the partial transposition of the post-measurement 
states to be larger than $\sim 10^{-4}$.

\end{document}